\documentclass[reprint, aps, prx, floatfix,superscriptaddress]{revtex4-2}

\usepackage{graphicx}    
\usepackage{dcolumn}    
\usepackage{bm}             
\usepackage{amssymb}   
\usepackage{amsmath}    
\usepackage{amsthm}
\usepackage{mathrsfs}    
\usepackage{commath}   
\usepackage{subfigure}    
\usepackage{braket}
\usepackage{color}
\usepackage{siunitx}
\usepackage[colorlinks=true, allcolors=blue]{hyperref}
\usepackage{hyphenat}
\usepackage{footnote}
\usepackage{notes2bib}
\usepackage{xcolor}

\newcommand{\beq}{\begin{equation}}
\newcommand{\eeq}{\end{equation}}

\begin{document}

\title{Microwave-Optical Entanglement from Pulse-pumped Electro-optomechanics}

\author{Changchun Zhong}
\email{zhong.changchun@xjtu.edu.cn}
\affiliation{Department of Physics, Xi'an Jiaotong University, Xi'an, Shanxi 710049, China}

\author{Fangxin Li}
\affiliation{Department of Physics, University of Chicago, Chicago, IL 60637, USA}

\author{Srujan Meesala}
\affiliation{Institute for Quantum Information and Matter, California Institute of Technology, Pasadena, California 91125, USA}

\author{Steven~Wood}
\affiliation{Kavli Nanoscience Institute and Thomas J. Watson, Sr., Laboratory of Applied Physics, California Institute of Technology, Pasadena, California 91125, USA}
\affiliation{Institute for Quantum Information and Matter, California Institute of Technology, Pasadena, California 91125, USA}

\author{David Lake}
\affiliation{Kavli Nanoscience Institute and Thomas J. Watson, Sr., Laboratory of Applied Physics, California Institute of Technology, Pasadena, California 91125, USA}
\affiliation{Institute for Quantum Information and Matter, California Institute of Technology, Pasadena, California 91125, USA}

\author{Oskar~Painter}
\affiliation{Kavli Nanoscience Institute and Thomas J. Watson, Sr., Laboratory of Applied Physics, California Institute of Technology, Pasadena, California 91125, USA}
\affiliation{Institute for Quantum Information and Matter, California Institute of Technology, Pasadena, California 91125, USA}
\affiliation{Center for Quantum Computing, Amazon Web Services, Pasadena, California 91125, USA}

\author{Liang Jiang}
\email{liang.jiang@uchicago.edu}
\affiliation{Pritzker School of Molecular Engineering, University of Chicago, Chicago, IL 60637, USA}

\date{\today}

\begin{abstract}

Entangling microwave and optical photons is one of the promising ways to realize quantum transduction through quantum teleportation. This paper investigates the entanglement of microwave-optical photon pairs generated from an electro-optomechanical system driven by a blue-detuned pulsed Gaussian pump. The photon pairs are obtained through weak parametric-down-conversion, and their temporal correlation is revealed by the second-order correlation function. We then study the discrete variable entanglement encoded in the time bin degree of freedom, where entanglement is identified by Bell inequality violation. Furthermore, we estimate the laser-induced heating and show that the pulse-pumped system features lower heating effects while maintaining a reasonable coincidence photon counting rate.

\end{abstract}

\maketitle

\section{Introduction}

Efficiently converting quantum information between different frequencies is crucial for scaling up distributed quantum architecture and advancing modern quantum networks \cite{cirac1997,kimble2008}. Recently developing quantum computation and communication modules utilize diverse physical platforms operating at varying energy scales, such as superconducting circuits in the microwave regime and communicating photons in the optical frequency range. A quantum transducer serves as a device designed to convert quantum information between microwave and optical photons \cite{lauk2020,han2021}, enabling the coherent connection between these two distinct modules.

The significant energy gap between microwave and optical photons presents substantial technological challenges in terms of material non-linearity and system noise when converting quantum information between them. In recent decades, various physical platforms, including electro-optomechanics \cite{Andrews2014,Regal2011,Bochmann2013,Taylor2011,Barzanjeh2011,Wang2012,Tian2010,*Tian2012,*Tian2014,Zou2016,Midolo2018,Bagci2014,Vainsencher16,Winger2011,Pitanti2015,mayor2024,zhao2024}, electro-optics \cite{Tsang2010,*Tsang2011,Javerzac-Galy2016,Fan2018,zhong2022}, magnons \cite{Hisatomi2016}, Rydberg atoms \cite{Hafezi2012,Kiffner2016,Gard2017}, and others, are under active investigation in order to achieve the goal of quantum transduction \cite{lauk2020,han2021}. Theoretically, quantum transducers for different frequency modes can be constructed using either beam splitter coupling for direct quantum conversion, known as direct quantum transduction (DQT), or two-mode squeezing interaction for generating microwave-optical entanglement, termed entanglement-based quantum transduction (EQT). EQT is expected to be more practical than DQT and has demonstrated compatibility with current technology \cite{zhong2020}. Indeed, compared to DQT, EQT offers a larger quantum capacity region for quantum information conversion within a more extensive practical parameter space \cite{zhong2022PRAPP,wu2021}. Furthermore, microwave-optical entanglement is compatible with the renowned DLCZ protocol, which holds the potential to directly link microwave quantum circuits \cite{zhong2020A,stefan2021}.

An essential step of EQT is the efficient generation of microwave-optical entanglement \cite{zhong2020}. Prior analyses of entanglement generation have primarily centered on the model involving the steady state generation through two-mode squeezing interactions with a continuous laser drive. This study, however, delves into a scenario where a piezo-optomechanical system is driven by short pump pulses which generally does not have sufficient time to reach a steady state. In comparison to continuous pumping, the pulsed drive model describes more accurately the prevalent time bin entanglement encoding scheme. Additionally, the system with a shorter pump suffers less on the laser induced heating, e.g., in chip-scale quantum transducers. This potentially allows for higher instantaneous laser pump power, increasing the probability of photon pair generation while maintaining quantum coherence.  

In the subsequent sections, we begin by analyzing a model system for quantum transduction based on piezo-optomechanics, where a blue-detuned Gaussian pump pulse is applied. The system's Hamiltonian features a time-dependent squeezing strength with a Gaussian time profile. We then investigate the temporal correlation of the resulting microwave-optical emission. With an overall weak laser pump, pair photon generation is the most probable outcome (excluding vacuum). The pair photon state and its temporal correlation can be described through the bi-photon wave packet. Employing the Schmidt decomposition, we further unveil the temporal mode structure of the wave packets, wherein the fundamental zeroth mode dominates the probability distribution. This output temporal mode can effectively encode discrete variable entanglement in time bin degrees of freedom. We numerically verify this time bin entanglement through Bell inequality violation using the state-of-the-art experimental parameters. Finally, we study a model for examining the laser-induced heating in the system, comparing both continuous laser pumps and pulsed drives. Our findings demonstrate that pulsed pumping allows for a significantly higher pump power while maintaining the same average heating noise as the continuous pump scheme. In practice, careful optimization of the duration and strength of the pump pulses is essential to suppressing laser-induced heating, which ensures the quantum coherence, and achieving a reasonable pair generation rate for practical applications in quantum networking.

\section{The model system}

We consider a piezo-optomechanical system with a short pump pulse. This system involves two main interactions: piezo-electricity induces a microwave-mechanical beam-splitter type coupling; the mechanical mode simultaneously interacts with optical photons through optical scattering forces \cite{han2020ca,mirhosseini2020}. With the pump pulse blue detuned, the scattering interaction usually reduces to a two-mode-squeezing coupling. We denote $\hat{a}$, $\hat{b}$ and $\hat{c}$ as the optical, mechanical and electrical mode operators, 
$\omega_o$, $\omega_m$ and $\omega_e$ as the corresponding mode frequencies, respectively. The total Hamiltonian is expressed as follows    
\begin{equation}
\begin{split}
\hat{H}/\hbar=&\omega_o\hat{a}^\dagger\hat{a}+\omega_m\hat{b}^\dagger\hat{b}+\omega_e\hat{c}^\dagger\hat{c}-g_{em}(\hat{b}^\dagger\hat{c}+\hat{b}\hat{c}^\dagger)\\
&-g_0\sqrt{\bar{n}_o(t)}(\hat{a}^\dagger\hat{b}^\dagger+\hat{a}\hat{b}).
\end{split}
\end{equation}
Here $g_{em}$ is the piezo-mechanical coupling and $g_0$ represents the one-photon optomechanical coupling rate. The latter is further modified by the intracavity photon number, denoted as $\bar{n}_o(t)$. We define $g_{om}(t):= g_0\sqrt{\bar{n}_o(t)}$ as the optical-microwave squeezing strength. It's important to note that the photon number changes with time due to the short pump pulse, resulting in a time-dependent squeezing strength. In this paper, we consider a pump pulse with a Gaussian time profile. As the optical cavity has a relatively large $\kappa_o$ over the optomechanical coupling rate, which essentially puts us in the adiabatic regime, the intracavity photon $\bar{n}_o(t)$ closely follows the pump pulse time profile, exhibiting a Gaussian time dependence. 

\section{The bi-photon output and the Schmidt modes}

\subsection{General theory}

The piezo-optomechanical system is capable of generating entangled optical-microwave photons. Intuitively, the two-mode squeezing interaction first entangles the optical and the mechanical modes, meanwhile the mechanical excitation is swapped to the electrical mode with a separate beam-splitter type coupling. The pair-photon generated from this process can be approximately described by the bi-photon wave packet \cite{brecht2015}
\begin{equation}
    \ket{\psi}=\iint dt_1dt_2f(t_1,t_2)\hat{a}^\dagger(t_1)\hat{c}^\dagger(t_2)\ket{0}.
\end{equation}
The coefficient $f(t_1,t_2)$ is determined by the two-time correlation function, which is given by
\begin{equation}\label{eq_co}   \abs{f(t_1,t_2)}^2\propto
\begin{cases} &\braket{\hat{a}^\dagger(t_1)\hat{c}^\dagger(t_2)\hat{c}(t_2)\hat{a}(t_1)}, t_1< t_2  \\
     & \braket{\hat{c}^\dagger(t_2)\hat{a}^\dagger(t_1)\hat{a}(t_1)\hat{c}(t_2)},t_2< t_1
\end{cases}.     
\end{equation}
The bi-photon wave packet can be uniquely decomposed into orthogonal temporal modes through the Schmidt decomposition \cite{nielsen2010} $f(t_1,t_2)=\sum^\infty_{k=0}\sqrt{\lambda_k}f^o_k(t_1)f^e_k(t_2)$. Thus, the wave packet can be rewritten as 
\begin{equation}
    \ket{\psi}=\sum_{k=0}^\infty\sqrt{\lambda_k}\ket{\psi_k^o}\ket{\psi_k^e},
\end{equation}
where the states $\ket{\psi_k^o}=\int dt_1f^o_k(t_1)\hat{a}^\dagger(t_1)\ket{0}$ and $\ket{\psi_k^{e}}=\int dt_2 f^e_k(t_2)\hat{c}^\dagger(t_2)\ket{0}$ are in the Hilbert space associated with the optical and microwave temporal modes, respectively. Pairs of temporal modes are excited with probability $\lambda_k$. In temporal mode degrees of freedom, the output is entangled with entanglement entropy $S=-\sum_k\lambda_k\ln\lambda_k$. Methods for efficiently generating and manipulating them is a current topic of interest in quantum information processing \cite{raymer2020}.

\subsection{Pair photon generated from Gaussian pump pulse}

\begin{figure}[t]
\centering
\includegraphics[width=\columnwidth]{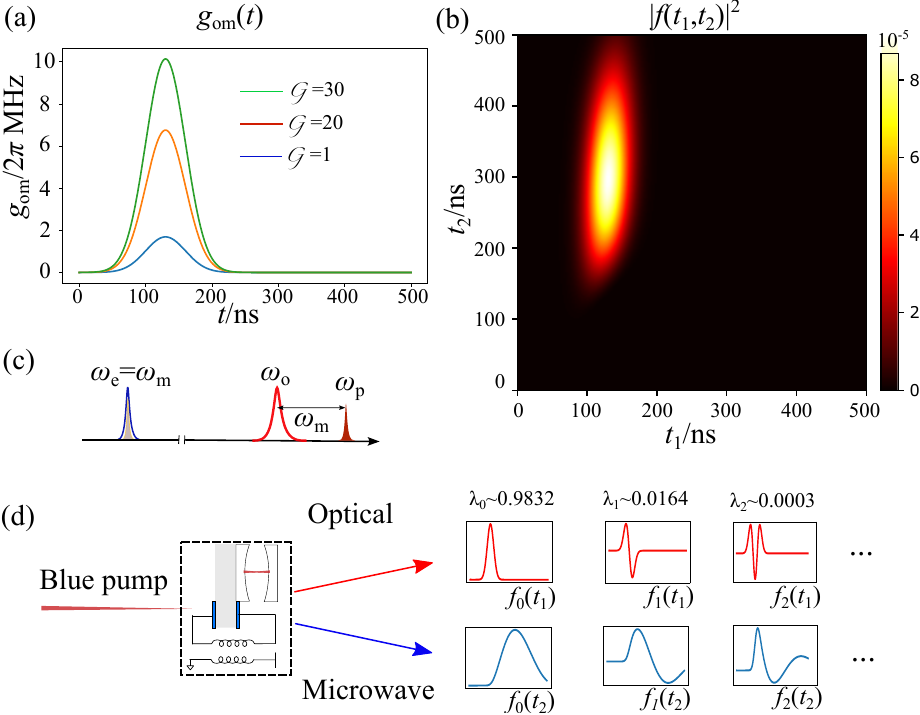}
\caption{(a) The time dependent squeezing strength $g_{om}(t)$ that takes a Gaussian form with varied $\mathcal{G}$. (b) The normalized amplitude of the output bi-photon wave packet for $\mathcal{G}=1$. (c) The frequency landscape for the piezo-optomechanical system under a blue detuned pump. (d) The decomposed temporal modes of the output wave packet with $\mathcal{G}=1$. \label{fig1}}
\end{figure}

We assume a pump pulse that generates a time dependent optomechanical squeezing strength
\begin{equation}
    g_{om}(t)= \mathcal{G}e^{-\frac{(t-t_0)^2}{2\sigma^2}}*g_0.
\end{equation}
For demonstration, we pick $\mathcal{G}=\{1,20,30\}$, $t_0=130$ ns and $\sigma=30$ ns for the pump, as shown in Fig.~\ref{fig1}(a). The frequency landscape of the system and the pump is given in Fig.~\ref{fig1}(c). Using the experimentally feasible parameters from Tab.~\ref{tab1}, we calculate the system dynamics by numerically solving the Lindblad master equation with QuTip software package \cite{qutip2012}
\begin{equation} \label{mast}
    \dot{\rho}=-\frac{i}{\hbar}[\hat{H},\rho]+\sum_{i=1}^{6}\frac{1}{2}\left(2\hat{L}_i\rho\hat{L}_i^\dagger-\{\hat{L}_i^\dagger\hat{L}_i,\rho\}\right),
\end{equation}
where $\rho(t)$ denotes the system density operator including the optical, mechanical and electrical modes.The jump operators are $\hat{L}_1=\sqrt{\kappa_o}\hat{a}$, $\hat{L}_2=\sqrt{\kappa_m(1+n_{th,b})}\hat{b}$, $\hat{L}_3=\sqrt{\kappa_m*n_{th,b}}\hat{b}^\dagger$, $\hat{L}_4=\sqrt{\kappa_{e,c}}\hat{c}$, $\hat{L}_5=\sqrt{\kappa_{e,i}(1+n_{th,c})}\hat{c}$ and $\hat{L}_6=\sqrt{\kappa_{e,i}*n_{th,c}}\hat{c}^\dagger$, where $\kappa_{o}=\kappa_{o,c}+\kappa_{o,i}$, $\kappa_m$ and $\kappa_e=\kappa_{e,c}+\kappa_{e,i}$ are the total dissipation rates for optical, mechanical and electrical mode, respectively. The sub-indexes ``c" and ``i" of the dissipation rates denote the corresponding out-coupling and intrinsic loss channels of a mode. $n_{th,b}$ ({$n_{th,c}$}) is the average thermal photon of the bath, which couples to the mechanical (electrical) mode intrinsically. Note in experiment, the system is usually placed in a several mK fridge, yielding a negligible thermal noise for the mechanical or electrical modes with the mode frequency on the order of several GHz. Thus, without specified otherwise we first take $n_{th,b}=n_{th,c}=0$ in the following discussions. More practical case with heating noise will be included in the end.

We first investigate the optical-microwave output state correlation, which is determined by its second order correlation function, as given by the amplitude of the bi-photon wave packet in Eq.~\ref{eq_co}. This correlation function with different time delay can be obtained with the help of the renowned quantum regression theorem \cite{gardiner2004}. Figure~\ref{fig1}(b) shows the result, where the shining blob appearing in the upper left corner indicates the probability in generating paired photons. Obviously, there is a tendency in detecting a microwave photon approximately $150$ ns after detecting an optical photon, which matches the expectation that it takes time for the mechanical excitation to be swapped to the microwave mode. 

Figure~\ref{fig1}(d) schematically shows the first several Schmidt decomposed temporal modes of the output bi-photon wave packet. It is noted that the output state mainly occupies the temporal zero mode, while other higher order modes are negligibly small. This is consistent with the pump which is also Gaussian in time. In the following, we will focus on the temporal zero mode, discuss the corresponding pair photon generation rate and the time bin encoding of the output entanglement.

\begin{table*}[t]
\caption{The following experimentally feasible parameters are used in the numerical evaluations in the text (unless specified otherwise).} \label{tab1}
\begin{center}
\begin{tabular}{c|c|c|c|c|c|c|c|c|c|c|c|c|c|c|c}
\hline
\hline
$g_{\text{em}}$/MHz  &  $\kappa_ {\text{e,i}}$/MHz & $\kappa_ {\text{e,c}}$/MHz & $\kappa_{\text{o,i}}$/(GHz)   & $\kappa_{\text{o,c}}$  & $\kappa_{\text{m}}$/kHz & $g_0$/kHz & $\omega_o$/THz & $\omega_m$/GHz & $\omega_e$/GHz & $\eta_o$ & $\eta_e$ & $T_o$ & $T_e$ & $D_o$/Hz & $D_e$/Hz \\
\hline
$2\pi\times 1.2$   & $2\pi\times 0.55$   & $2\pi\times 1.25$ &  $2\pi\times 0.65$   & $\kappa_{\text{o,i}}$  &  $2\pi\times150$ & $2\pi\times 260$ & $2\pi\times 190$ & $2\pi\times 5$ & $2\pi\times 5$ & 0.8 & 0.9 & $10^{-2}$ & $0.5$ &  $\sim 10$  & $10^3$ \\
\hline
\hline
\end{tabular}
\end{center}
\end{table*}

\section{The M-O photon pair generation and the time bin Bell state}

\subsection{The photon pair generation rate}

The photon pair generation rate is an important quantity in entanglement-based quantum transduction \cite{zhong2020}. We can theoretically calculate it in the framework of a second order correlation function \cite{glauber1963}. Consider at time $t$ during the pump, the coincidence click probability can be evaluated according to the formula
\begin{equation}
P(t,\tau)=G_2(t,\tau)*t_w^2,
\end{equation}
where $G_2(t,\tau)=\braket{\hat{a}_{out}^\dagger(t)\hat{c}_{out}^\dagger(t+\tau)\hat{c}_{out}(t+\tau)\hat{a}_{out}(t)}$ is the second order correlation function defined on the output modes. Note we have the input-output relation $\hat{a}_{out}=\hat{a}_{in}+\sqrt{\kappa_{o,c}}\hat{a}$ and $\hat{c}_{out}=\hat{c}_{in}+\sqrt{\kappa_{e,c}}\hat{c}$. $a_{in}$ and $c_{in}$ are the input bath-mode operators which are assumed to be vacuum unless specified otherwise. $t_w$ is the detection time window and $\tau$ is the time delay, which essentially means the time difference for recording an optical photon and subsequently a microwave photon. Denoting $r_D$ as the experiment repetition rate, then the coincidence counting rate within $t_w$ is given as 
\begin{equation}
R_{cc}(t,\tau)=P(t,\tau)*r_D.
\end{equation}
Using the quantum moment factoring theorem and the fact $r_D*t_w\sim 1$, we get
\begin{equation}\label{crate}
\begin{split}
    R_\text{cc}(t,\tau)\simeq R_a+R_c,
\end{split}    
\end{equation}
where $R_a=\braket{a^\dagger_\text{out}(t)a_\text{out}(t) }\braket{c^\dagger_\text{out}(t+\tau)c_\text{out}(t+\tau)}*t_w$ is called the accidental counting rate while $R_c=\braket{a^\dagger_\text{out}(t)c^\dagger_\text{out}(t+\tau) }\braket{c_\text{out}(t+\tau)a_\text{out}(t)}*t_w$ is the correlated counting rate. Note Eq.~\ref{crate} gives the coincidence photon counting rate at time $t$ after the pump, and the total rate can be obtained by taking the sum over the entire pump duration 
\begin{equation}
\begin{split}
    R_\text{cc}(\tau)=&\sum_{k=0}^N G_2(t_k,\tau)*t_w, \text{ with $t_k=t_i+k*t_w$}\\
    \simeq&\int^{t_{f}}_{t_{i}} G_2(t,\tau)dt,
\end{split}
\end{equation}
where $t_{i}$ ($t_f$) denotes the initial (final) time of the pump duration and $t_f=t_i+N*t_w$. The second line of integration takes into account that the time $t_w$ is much smaller than the output photon line-width. Obviously, the coincidence rate is $\tau$-dependent. As shown in Fig.~\ref{fig2}(a) by numerical calculation (details given in the following sections), the rate is maximized with time delay $\tau\simeq 150$ ns, which matches the expectation that it takes time ($\sim\kappa_e/g^2_{em}$) to swap the mechanical mode excitation to the microwave mode. 


In practice, the coincidence counting rate is also affected by the photon transmission loss, detector dark count (including pump leakage) and detector efficiency. The pump leakage induced dark count is usually proportional to the pump power. Denote $T_o,T_e$ as the transmission coefficients, $D_o,D_e$ as the dark count rates and $\eta_o,\eta_e$ the detector efficiencies for the optical and microwave photons, respectively. The coincidence rate Eq.~\ref{crate} shall be modified as
\begin{equation}
    R_\text{cc}(t,\tau)\simeq R_a^\prime+R_c^\prime,
\end{equation}
where $R_a^\prime=(\eta_oT_o\bar{n}^o_\text{out}(t) +D_o)(\eta_eT_e*\bar{n}^e_\text{out}(t+\tau)+D_e)*t_w$ and $R_c^\prime=\eta_o\eta_eT_oT_eR_c*t_w$. we denote $\bar{n}^o_\text{out}(t)=\braket{a^\dagger_\text{out}(t)a_\text{out}(t)}$, and $\bar{n}^e_\text{out}(t+\tau)=\braket{c^\dagger_\text{out}(t+\tau)c_\text{out}(t+\tau)}$. The typical value of the parameters are given in Tab.~\ref{tab1}, and shall be used in the following numerical estimations.

\color{black}

\subsection{Time bin Bell state}

Using the system, it is convenient to generate Bell states in time bins, which is a coding scheme resistant to photon loss. In each experiment, we pump the system with two weak successive pulses separated in time by $\Delta t$, defining the two time bins. Due to the low pump power, one and only one pair of photons can be generated with high probability (excluding vacuum). We can make the photons in the two time bins indistinguishable by controlling the time delay on both microwave and optical sides, creating a Bell state 
\begin{equation}
    \ket{\psi}_B=\frac{1}{\sqrt{2}}\left(\ket{\text{bin}_1}_e\ket{\text{bin}_1}_o+\ket{\text{bin}_2}_e\ket{\text{bin}_2}_o\right),
\end{equation}
where we denote the optical time bin state $\ket{\text{bin}_i}_o=
\hat{a}_\text{out,i}^\dagger\ket{0}$ and the microwave time bin state $\ket{\text{bin}_i}_e=\hat{c}_\text{out,i}^\dagger\ket{0}$ with the time bin index $i=\{1,2\}$. The entanglement in this state can be verified using different entanglement witnesses, e.g., the Bell state fidelity \cite{meesala2023B}. In the following, we shall discuss its Bell inequality violation with different parameters and compare the corresponding photon generation rates.

\subsection{Entanglement verification}

\begin{figure*}[t]
\centering
\includegraphics[width=\textwidth]{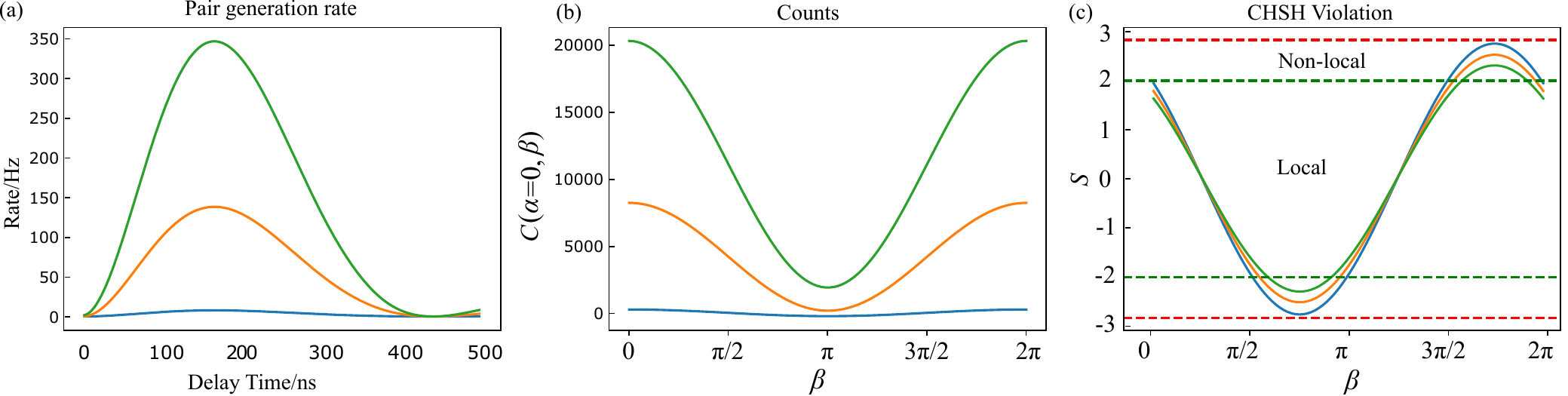}
\caption{(a) The optical-microwave photon generation rate in terms of the time delay. (b) The coincidence counts within 1 minutes with varied angle $\beta$ and fixed time delay $\tau=150$ ns. (c) The CHSH inequality violation curves. The dashed-green-horizontal lines give the local and non-local boundary while the red dashed line marks the maximal violation value $2\sqrt{2}$. The green, orange and blue curves in (a), (b) and (c) correspond to the pumps with $\mathcal{G}=\{30,20,1\}$.\label{fig2}}
\end{figure*}

In experiment, the entanglement can be tested by counting the correlated statistics with proper control and photon detection. For instance, on the optical side, a Mach-Zehnder interferometer can be used rotate the optical time bin in the Bloch sphere. If we fix the qubits in the equator, the detection of a single optical photon at the interferometer output will project the state onto 
\begin{equation}
\ket{\psi^\pm_o}=1/\sqrt{2}(\ket{\text{bin}_1}_o\pm e^{i\theta}\ket{\text{bin}_2}_o).    
\end{equation} 
Note half of the photon rate will be lost since only delayed photon in time bin $1$ can interfere with non-delayed photon in time bin $2$. On the microwave side, the photon can first be converted to a transmon qubit excitation which effectively selects the state by later qubit rotation \cite{campagne2018,axline2018}
\begin{equation}
\ket{\psi^\pm_e}=1/\sqrt{2}(\ket{\text{bin}_1}_e\pm e^{i\phi}\ket{\text{bin}_2}_e). 
\end{equation}
The statistics $\braket{\hat{a}_\text{out}^\dagger(t)\hat{c}_\text{out}^\dagger(t+\tau)\hat{c}_\text{out}(t+\tau)\hat{a}_\text{out}(t)}$ with specific configurations, e.g., given $\theta,\phi$, can be obtained by repeating the experiment many times \cite{meesala2023}. To simplify the notation, we will omit the lower index ``out" unless required for clarity in the following discussions.

Theoretically, this coincidence rate can be straightforwardly calculated. In the Heisenberg picture, the optical-microwave output mode operators actually undergo the transforms
\begin{equation}\label{tta}
    \begin{pmatrix}
        a_1^\prime \\a_2^\prime
    \end{pmatrix}=
    \begin{pmatrix}
        \cos\alpha/2 & \sin\alpha/2 e^{i\theta} \\
        -\sin\alpha/2 e^{-i\theta} & \cos\alpha/2
    \end{pmatrix}
    \begin{pmatrix}
        a_1 \\ a_2
    \end{pmatrix}
\end{equation}
and
\begin{equation}\label{ttb}
    \begin{pmatrix}
        c_1^\prime \\c_2^\prime
    \end{pmatrix}=
    \begin{pmatrix}
        \cos\beta/2 & \sin\beta/2 e^{i\phi} \\
        -\sin\beta/2 e^{-i\phi} & \cos\beta/2
    \end{pmatrix}
    \begin{pmatrix}
        c_1 \\ c_2
    \end{pmatrix}
\end{equation}
corresponding to those time bin mode rotations in the whole range of the Bloch sphere. For any fixed $\alpha,\beta,\theta$ and $\phi$, one can record four different outcomes, e.g., simultaneously detecting the mode $a^\prime_1$ and $c_1^\prime$, which as we discussed before has the coincidence rate (take $\theta=0,\phi=0$ for instance)
\begin{equation}\label{Rcc}
\begin{split}
    &\int_{t_i}^{t_f}\braket{a^{\prime\dagger}_1(t)c^{\prime\dagger}_1(t+\tau)c^\prime_1(t+\tau)a^\prime_1(t)}dt\\         =&\int_{t_i}^{t_f}\braket{a^{\prime\dagger}_1(t)a^\prime_1(t)}\braket{c^{\prime\dagger}_1(t+\tau)c^\prime_1(t+\tau)}\\ &+\braket{a^{\prime\dagger}_1(t)c^{\prime\dagger}_1(t+\tau)}\braket{c^{\prime}_1(t+\tau)a^{\prime}_1(t)}dt\\     =&\int_{t_i}^{t_f}\braket{a_1^\dagger(t)a_1(t)}\braket{c_1^\dagger(t+\tau)c_1(t+\tau)}+\\ &\cos^2\frac{\alpha+\beta}{2}\braket{a_1^\dagger(t)c_1^\dagger(t+\tau)}\braket{c_1(t+\tau)a_1(t)}dt,
\end{split} 
\end{equation}
where the first equal sign is arrived at by the renowned moment factoring theorem. The last equal sign is obtained using Eq.~\ref{tta}$-$\ref{ttb}, and we use the fact that the optical (microwave) photons in time bin 1 are not correlated with photons in time bin 2, e.g., $\braket{\hat{a}^\dagger_1(t)\hat{c}^\dagger_2(t+\tau)}=0$. Further, suppose we collect data for a certain amount of time, e.g., $t_\text{c}$, we can get the corresponding coincidence counts, denoted as
\begin{equation}\label{counts}
    C_{(\alpha,\beta)}=R_\text{cc}(\hat{a}_1^\prime,\hat{c}_1^\prime)\times t_\text{c}.
\end{equation}  

\subsection{CHSH inequality}

The well-known CHSH inequality is a type of Bell inequality, whose violation excludes the possibility of local hidden variable theory. It is written as $\abs{S}\le 2$ with
\begin{equation}\label{chsh}   S=\braket{\sigma_A\sigma_B}+\braket{\sigma_{A^\prime}\sigma_{B^\prime}}+\braket{\sigma_A\sigma_{B^\prime}}-\braket{\sigma_{A^\prime}\sigma_B},
\end{equation}
where $\sigma_A$ is the Pauli-Z operator along $A$ direction, e.g., $\sigma_A=\ket{\alpha,\theta}\bra{\alpha,\theta}-\ket{\alpha+\pi,\theta+\pi}\bra{\alpha+\pi,\theta+\pi}$ for $A=(\alpha,\theta)$ in the Bloch sphere of the optical time bin. Similarly, we have $\sigma_B=\ket{\beta,\phi}\bra{\beta,\phi}-\ket{\beta+\pi,\phi+\pi}\bra{\beta+\pi,\phi+\pi}$ for $B=(\beta,\phi)$ of the microwave time bin. The term, e.g., like $\ket{\alpha,\theta}\bra{\alpha,\theta}$ is called an ``event" in quantum information theory \cite{busch2016}, which corresponds to the photon click that projects the photon state onto, e.g., $\ket{\psi}=\cos\alpha/2\ket{\text{bin}_1}_o+e^{i\theta}\sin\alpha/2\ket{\text{bin}_2}_o$. The corresponding number of clicks in a given amount of time equals exactly the term $C_{(\alpha,\beta)}$ that we have in Eq.~\ref{counts} (with phases included). The other terms in Eq.~\ref{chsh} are similar. Typically, the measurement setting for violating CHSH is fixed in a plane. Without losing generality, we choose the plane with $\theta=\phi=0$, and $\alpha^\prime=\alpha+\pi/2,\beta^\prime=\beta+\pi/2$. Thus, $S$ becomes
\begin{equation}   S(\alpha,\beta)=\braket{\sigma_\alpha\sigma_\beta}+\braket{\sigma_{\alpha^\prime}\sigma_{\beta^\prime}}+\braket{\sigma_\alpha\sigma_{\beta^\prime}}-\braket{\sigma_{\alpha^\prime}\sigma_\beta},
\end{equation}
with each term given by, e.g.,
\begin{equation}    \braket{\sigma_\alpha\sigma_\beta}=\frac{C_{(\alpha,\beta)}+C_{(\alpha+\pi,\beta+\pi)}-C_{(\alpha,\beta+\pi)}-C_{(\alpha+\pi,\beta)}}{C_{(\alpha,\beta)}+C_{(\alpha+\pi,\beta+\pi)}+C_{(\alpha,\beta+\pi)}+C_{(\alpha+\pi,\beta)}}.
\end{equation}
Combining the above equations with Eq.~\ref{counts}, the quantity $S$ can be readily evaluated.

\subsection{Numerical results}

We first numerically calculate the coincidence rate from Eq.~\ref{Rcc} in terms of the time delay by further fixing $\alpha=\beta=0$, as shown in Fig.~\ref{fig2}(a). Given three different model pumps, the coincidence rate increases with the laser pump strength. Figure~\ref{fig2}(b) depicts the coincidence counts in $1$ minute while varying the angle $\beta$. Although a larger pump gives higher coincidence counts, it also increases the accidental counts which could potentially ruin the Bell test. Intuitively, it is known that larger pump power could raise the chance of higher-order photon excitation, which degrades the Bell state fidelity. This is confirmed by calculating the quantity $S$ as a function of $\beta$. The result is shown in Fig.~\ref{fig2}(c). All three curves go outside the region $\abs{S}\le 2$, clearly violating the CHSH inequality. Also, the larger violation (blue curve) is obtained from lower laser pump power, signifying the tradeoff between Bell state fidelity and the photon pair generation rate. In practice, one would optimize for higher fidelity or higher photon generation rate depending on the specific application or requirements.

\section{Laser heating estimation}

\begin{figure*}[t]
\centering
\includegraphics[width=\textwidth]{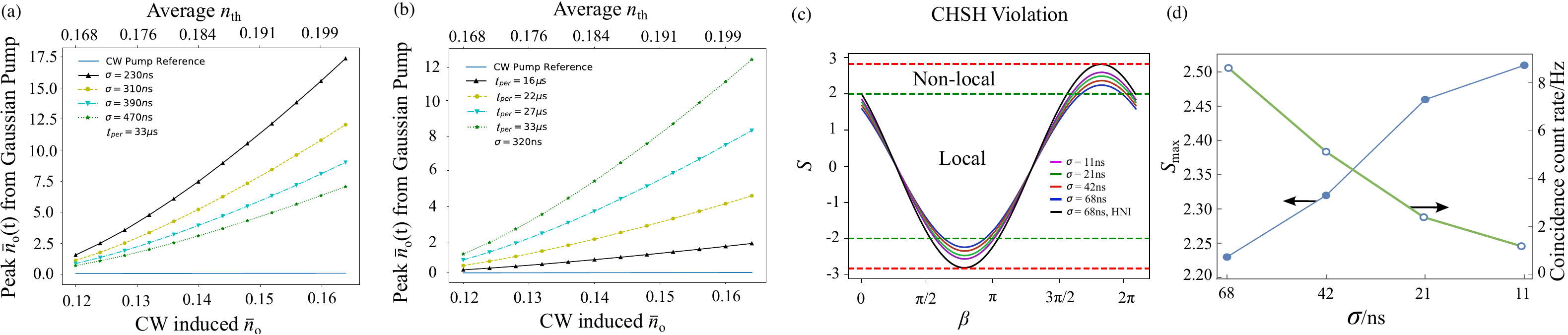}
\caption{(a) and (b) Comparison of the intensity of the CW pump and the peak intensity of the Gaussian pump that generates the same average thermal occupation over time. The top $x$-axis shows the average thermal occupation averaged of time. (a) is for fixed repetition period $t_{\text{per}}=33\text{ }\mu$s, the peak pulse power for different pulse width $\sigma$. (b) is for fixed pulse width $\sigma=320$ ns, the peak pulse power for different pulse repetition period $t_{\text{per}}$. (c) The Bell inequality violation curves including time dependent heating noises by varying the pump width $\sigma$. As a contrast, the black curve shows the result without heating noise (In the simulation, all other parameters are kept the same). (d) The maximal violation $S_\text{max}$ and the coincidence rate for different pump widths corresponding to the data in (c). \label{fig3}}
\end{figure*}

\subsection{A simplified model}

Laser heating is one of major problems in coherently controlling the quantum transduction process. In experiment, one needs a dilution refrigerator to maintain a cryogenic environment free of thermal noise. If the control laser power used to activate the photon-pair generation of the transducer is too high, it can lead to excess heating within the transducer due to parasitic optical absorption, leading to loss of coherent transduction of signals. For a typical quantum transducer with a continuous-wave (CW) pump, the pump power must be severely limited to avoid heating effects, greatly limiting the mode squeezing strength and the instantaneous pair generation probability. However, given the transient heating of the transducer, in the case of a pulsed pump, the peak power can be much larger while still avoiding the deleterious effects of heating. This indicates that pulsed pumping, in addition to being part of our time bin entanglement approach, can also potentially yield a much higher instantaneous squeezing strength in the presence of parasitic heating effects than that of CW pumping. 



Here, we study the heating dynamics of the Gaussian pulse pump and compare it with that of the CW pump as a quantitatively guide to experiments. {For steady-state bath population,the heating dynamics in nanoscale piezo-optomechanical transducers is very well-described by the \textit{bottleneck} model in Ref.~\cite{oskar4}, which yields a prediction for both the transient behavior of the thermal bath occupancy of the microwave (mechanical) mode and its damping rate to the hot bath. Here we employ a much simpler model that captures the essential transient behavior in these devices. In the long time limit, it matches with the scaling prediction of the steady-state occupancy given in \cite{oskar4}. Roughly, the dynamics is governed by}
\begin{equation}\label{heat capacity}
    (P_\text{heat}-P_\text{cool})\times \Delta t = \mathcal{CM}\times\Delta T,
\end{equation}
where $P_\text{heat}$ and $P_\text{cool}$ denote the laser heating and system cooling power, respectively. {In this model, we assume the system is described by an effective temperature $T_p$ (temperature distribution across the device would be more practical and the discussion is put in the appendix).} $\Delta T$ gives the temperature change. $\Delta t$ is the time interval. $\mathcal{C}$ and $\mathcal{M}$ are the heat capacity and mass of the piezo-material. In practice, the cooling power and the heat capacity can be temperature dependent. Since we care about the regime where the temperature variation is small, it is reasonable to take the linear dependence of the cooling power on temperature. Assuming a constant heat capacity, we solve the time dependence of the device temperature with two different heating powers: 1) a constant $P_\text{heat}=h_0+h_1$; 2) a Gaussian heating power $P_\text{heat}=h_0+h_2*\exp(-\frac{(t-t_0)^2}{2\sigma^2})$, where $h_0$ represents the environment heating and the second term in each case is induced by the laser pump. For a complete comparison with CW pump, we also consider applying the Gaussian pump repeatedly with the periodicity $t_\text{per}$.  

The results are shown in Fig.~\ref{fig3} (the detailed calculations are put in Appendix A). For a CW pump and a Gaussian pulse pump that induces the same average thermal noise, the allowed instantaneous intra-cavity photon of the pulsed pump can be much larger than that of the CW pump. The data in Fig.~\ref{fig3}(a) and \ref{fig3}(b) confirms this intuition for various pulse period $t_\text{per}$ and the width $\sigma$ of the Gaussian pump. The larger intra-cavity photon indicates a bigger coupling strength, which suggests a higher probability in generating paired photons while the average added thermal noise stays the same.  




\subsection{The Bell test with heating noise}

The above model enables us to simulate the entangled photon source with laser heating noise. We consider laser heating that generates an effective time dependent thermal noise $n_\text{th}(t)$ that couples to the mechanical mode. As shown in the appendix, we fit our model to the experimental data \cite{meesala2023}, and obtain a time-dependent thermal noise 
\begin{equation}
    n_\text{th}(t)=(e^{{\hbar\omega_m}/{k_bT_p(t)}}-1)^{-1}.
\end{equation}
The thermal temperature is given by
\begin{equation}
    T_p(t)=b^\prime e^{-at}f(t)+\frac{d}{a}[1-(1-\frac{a}{d}T_0)e^{-at}],
\end{equation}
where the function $f(t)=0.305-0.306*\text{Erf}(2.013-10.408t)$. All other parameters are given in the appendix. The pump laser also generates the intracavity photon
\begin{equation}
    \bar{n}_o(t)=n_m*e^{\frac{(t-t_0)^2}{2\sigma^2}}
\end{equation}
where $n_m=0.8$, $t_0=160$ ns and $\sigma=68$ ns, which further determines the optomechanical squeezing strength. Using the above new parameters, we numerically solve the master equation \ref{mast} again and calculate the Bell inequality violation curve. The results are shown in Fig.~\ref{fig3}(c). The blue curve clearly shows that the system is producing entangled pairs beyond any classical descriptions, which matches the claim in \cite{meesala2023B} based on non-classical values of entanglement fidelity. As a prediction for the trend, we also perform the simulation with varied pump pulse widths $\sigma=\{42,21,11\}$ ns, while the other parameters are kept the same. It is shown that a larger violation can be achieved by decreasing the pump pulse width. Figure 3(d) compares the trade-off between obtainable entanglement fidelity and rate for the different pump pulse widths.

\color{black}

\section{Discussion}

The quantum transducer is an essential device in the development of distributed quantum architectures. Recent experimental demonstrations of microwave-optical entanglement, encoded either in the continuous variable or discrete variable degrees of freedom \cite{sahu2023,meesala2023B}, have shown great promise in the realization of entanglement-based quantum transduction \cite{zhong2020}. Here we present a systematic study of a pulsed entanglement-based scheme, utilizing time bin degrees of freedom for encoding. Our model serves as a useful framework to analyze various physical platforms; however, in order to highlight practical limitations, and trade-offs between photon pair generation rate and Bell state fidelity, we consider a state-of-the-art piezo-optomechanical transducer~\cite{han2020ca,meesala2023,mirhosseini2020}. Our results indicate that such entanglement-based quantum transduction protocols should be able to realize. In the future, it would be helpful to consider different protocols, such as a well-controlled pump pulse shape, to further optimize the entangled resources and to make full use of them in connecting different quantum modules. We leave that for future research.

\begin{acknowledgments}
C.Z. thanks Mankei Tsang for helpful discussions. We acknowledge supports from 
the ARO(W911NF-18-1-0020, W911NF-23-1-0077), ARO MURI (W911NF-21-1-0325), AFOSR MURI (FA9550-19-1-0399, FA9550-21-1-0209, FA9550-23-1-0338), DARPA (HR0011-24-9-0359, HR0011-24-9-0361), NSF (OMA-1936118, ERC-1941583, OMA-2137642, OSI-2326767, CCF-2312755), NTT Research, and the Packard Foundation (2020-71479). C.Z. thanks the start up support from Xi'an Jiaotong University (Grant No. 11301224010717).
\end{acknowledgments}

\bibliography{all}

\onecolumngrid

\begin{appendix}

\section{A Phenomenological Heating Model}\label{heating model}

Laser induced device heating creates an effective thermal bath. {The microscopic dynamics of the optically-induced bath is described in detail by the \textit{bottleneck} model in \cite{oskar4}. According to the model, the geometry of the nanobeam imposes an effective phonon bottleneck which prevents a rapid thermalization between the higher frequency phonon modes and low-lying modes. The absorbed optical power from the laser populates the higher frequency phonons, resulting a buildup in the bath phonon
population above the bottleneck. This elevated-temperature bath then couples to the lower-lying modes—in particular the breathing mode at 5 GHz—through elastic 3-phonon scattering. The model provides a experimentally-verified scaling behavior of the bath occupancy and damping rate assuming a steady state bath population created by the laser. In this paper, we provide a simple prediction model of the transient behavior of the thermal bath. 

For the transient behavior of the thermal bath in the pulse-pumped scheme we consider, as a first step, a heating model whose long-time behavior matches with the steady state prediction in \cite{oskar4}. The first ingredient of our model comes from \cite{oskar3}, where the temperature of the effective thermal bath is understood as the macroscopic mean temperature $T_p$ of device with heat source defined at the geometric center of the transduction device and the boundary cooled almost to zero Kelvin. In this way, we use the macroscopic thermodynamics to describe the heating of the device. The second ingredient comes from the sub-linear scaling relation predicted by the \textit{bottleneck} model in \cite{oskar4}. In the long-time limit when the bath reaches a steady state, the bath temperature $T_p$ should relate to the laser induced cavity photon number $\bar{n}_o$ by a power law $T_p = \eta*\bar{n}_o^\gamma$, where $\eta$ and $\gamma$ are some positive real numbers depending on the experimental setting\cite{oskar3,oskar1,oskar2,oskar4}. } 


The transient laser heating behavior can be described using heat capacity equation: 
\begin{equation}
    (P_\text{heat}-P_\text{cool})\times \Delta t = \mathcal{CM}\times\Delta T,
\end{equation}
where $P_\text{heat} = h_0 + h(t)$. $h(t)=h_1$ for CW pump, and $h(t)=h_2*e^{-\frac{(t-t_0)^2}{2\sigma^2}}$ for a Gaussian pulse pump. In principle, $P_\text{cool}$ could be a general function of temperature $T$, but assuming that the temperature variation during the transient heating is sufficiently small, we can linearize the function and write, without loss of generality, $P_\text{cool}(T) = \alpha T + \beta $. Generally, the specific heat $\mathcal{C}$ is also temperature dependent but we model it as a constant since the temperature variation is small. Taking the time interval $\Delta t$ to be infinitesimal, we obtain the differential equation for the quasi-static thermal process: 
\begin{equation}\label{heatingeqn}
    \frac{dT}{dt} = \frac{h_0 + h(t)-\alpha T - \beta}{\mathcal{CM}}.
\end{equation}
This equation can be solved analytically. For CW pump, the solution is 
\begin{equation}\label{CWheat}
    T_{\text{CW}}(t)=\frac{b}{a}\left(1-\left(1-\frac{T_0a}{b}\right)e^{-at}\right),
\end{equation}
where the coefficients $a=\alpha/\mathcal{CM}$, $b=(h_1+h_0-\beta)/\mathcal{CM}$, and $T_0$ is the initial temperature at $t=0$. For the given set of parameters, the temperature of the chip with CW pump will equilibrate to $b/a$. 

It is worth mentioning that this solution also contains the solution to heating dynamics of the square wave pulse pump. When the pulse is on, the temperature would increase as in Eq.~\ref{CWheat}, and when the power is turned off, the device would also cool according to the same equation, but with the appropriate initial condition for continuity condition and $b$ replaced by $d = (h_0-\beta)/\mathcal{CM}$. Therefore, depending on the sign of the coefficient of $e^{-at}$, we could have heating with function of the form $u_1(1-v_1*e^{-at})$, and cooling with exponential decay $u_2*e^{-at}+v_2$, where $u_1,v_1,u_2,v_2$ are some arbitrary positive constants. Such heating and cooling behavior predicted by our simple model is validated by experimental observation in \cite{oskar1,oskar4}. 

For Gaussian pulse pump, the temperature solution is given by
\begin{equation}\label{gaussheat}
    T_{\text{gauss}}(t)= b^\prime*e^{-at} \int_0^t\,dt'\,\left(e^{at'}\,e^{-\frac{(t'-t_0)^2}{2\sigma^2}}\right) + \frac{d}{a}\left(1-\left(1-\frac{T_0a}{d}\right)e^{-at}\right),
\end{equation}
where $a=\alpha/\mathcal{CM}$, $b^\prime = h_2/\mathcal{CM}$, and $d=(h_0-\beta)/\mathcal{CM}$. We can quickly check that by replacing $e^{-\frac{(t'-t_0)^2}{2\sigma^2}}$ in the integral by $h_1/h_2$, we re-obtain the solution in Eq.~\ref{CWheat} knowing that $b=b^\prime h_1/h_2 + d$. The Gaussian integral in the first term of Eq.~\ref{gaussheat} can be rewritten as an error function, which would approach a constant value as $t\rightarrow \infty$. Hence, we know that after the Gaussian pulse, the system will be asymptotically cooled to $d/a$ following an exponential decay, which is the same cooling behavior as in the square pulse case. 

\begin{figure*}
\centering
\includegraphics[width=0.5\textwidth]{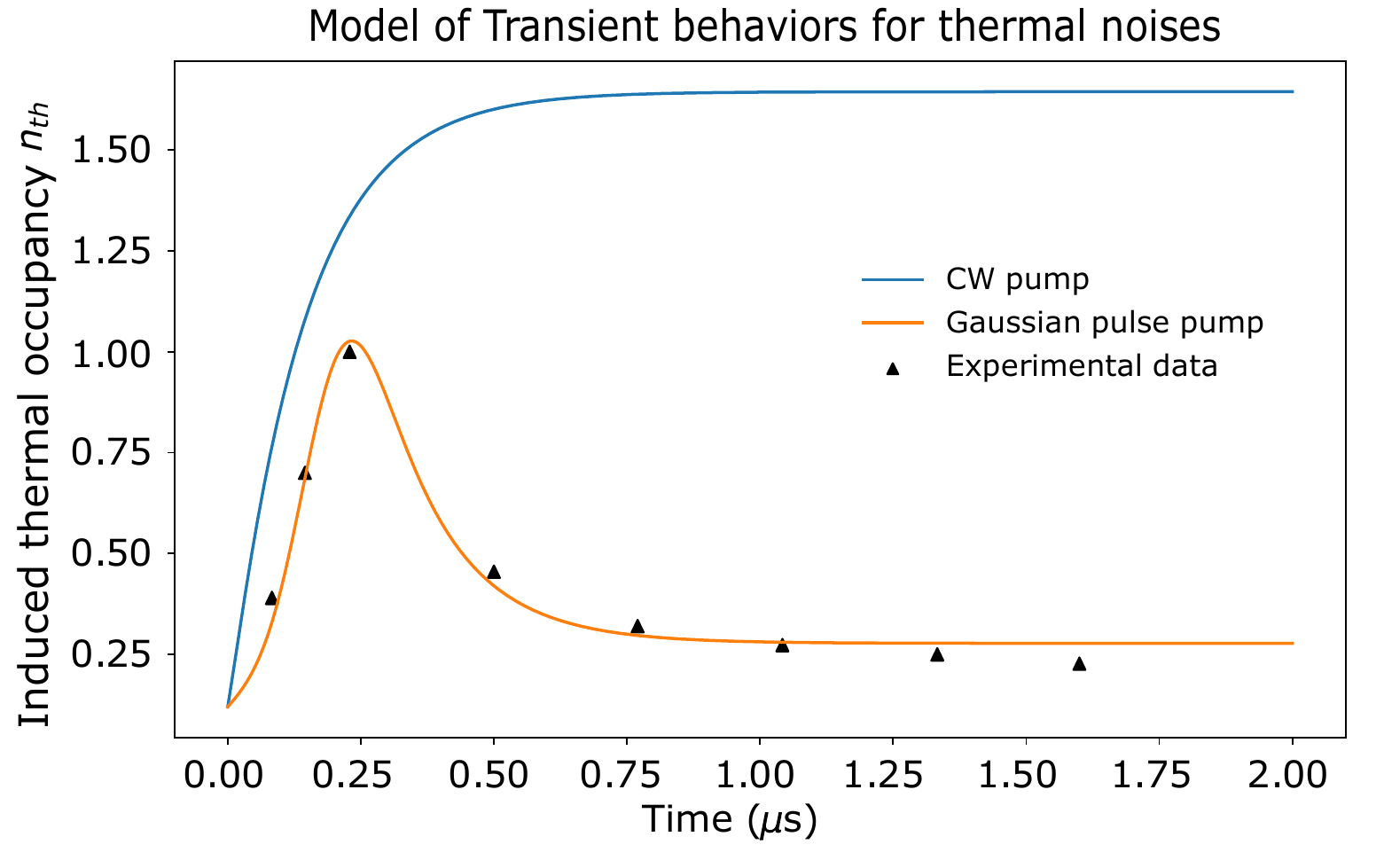}
\caption{Model of the transient heating behavior of the CW pump and the Gaussian pulse pump with parameters extracted from experimental data by fitting Eq.~\ref{gaussheat}.\label{heatmodel} Least squares fitting gives $a=7.244, b^\prime = 2.521, d=1.138, T_0=0.108$ K. $t_0=0.16\mu$s and $\sigma=0.068\mu$s are the Gaussian parameters obtained from the experiment. The CW pump would induce a higher thermal occupancy at thermal equilibrium.}
\end{figure*}

To make our model experimentally relevant, we fit the thermal bath dynamics measured in Ref.~\cite{meesala2023} with a function of the form in Eq.~\ref{gaussheat}. It turns out the model fits the experiment data remarkably well. The four fitting parameters revealed the system parameters $\{a,b^\prime,d, T_0 \}$, which we then use to infer the heating dynamics of the CW pump as in Eq.~\ref{CWheat}. The transient behavior for both the CW pulse and the Gaussian pulse are shown in Fig.~\ref{heatmodel}, which shows, as expected, the CW pump would induce a higher thermal occupancy at the thermal equilibrium. 

We have obtained the transient heating dynamics of the transducer given a fixed laser input $P_{\text{heat}}$. To see how the transient heating dynamics changes with laser intensity, we must set the equilibrium temperature of the CW pump to scale as power law with laser intensity: $ \eta * \bar{n}_o^\gamma=b/a =(b'h_1/h_2+d)/a$. If we assume that the system parameters $\{a,d, T_0\}$ does not change as we adjust the laser intensity and the instantaneous heating rate is proportional to the intro-cavity photon, then we can relate the parameters for the Gaussian pulse pump to the laser intensity by $b' h_1/h_2=a*\eta*\bar{n}_o^\gamma -d$. This would inform us how the temperature response $T_{\text{gauss}}(t)$ changes accordingly.  

To compare CW pump and Gaussian pump pulse, we assume that the pulse is repeated in a sequence with time period $t_{\text{per}}$, then the temperature response $ T^{\,\text{seq}}_{\text{gauss}} (t)$ is a piece-wise function. In each time period, it is $T_{\text{gauss}}(t)$, but with different initial conditions to match the continuity condition. 

To finally compare the heating of CW and Gaussian pulse, we want to find laser intensity $\bar{n}_o$ for the former and peak laser intensity $\bar{n}_o(t)$ for the latter such that the average photon occupancy during experiment is the same for both of them.
\begin{equation}
    \frac{1}{t_{\text{per}}}\int_{t_0}^{t_0+t_{\text{per}}} dt \,T^{\text{\,seq}}_{\text{gauss}} (t) = \frac{1}{t_{\text{per}}}\int_{t_0}^{t_0+t_{\text{per}}} dt \, T_{\text{CW}}(t),
\end{equation}
where $t_0$ is some time we picked in numerical calculation such that thermal equilibrium with the CW pump has almost been reached. The result of this comparison is shown in Fig.~\ref{fig3} in the main text.

\end{appendix}

\end{document}